\documentclass[traditabstract]{aa}
\usepackage{txfonts}
\usepackage{natbib,lineno} \bibpunct{(}{)}{;}{a}{}{,} 
\usepackage{color}
\usepackage[utf8]{inputenc}
\usepackage[T1]{fontenc}
\usepackage{graphicx}
\usepackage{float}
\usepackage{subfigure}
\begin{document}

\title{Making Faranoff-Riley I radio sources} \subtitle{II. The effects of jet magnetization}

\author{S. Massaglia\inst{1}, G. Bodo\inst{2}, P. Rossi\inst{2},
  S. Capetti\inst{2}, A. Mignone\inst{1}}

\authorrunning{S. Massaglia et al.}
\titlerunning{Making Faranoff-Riley I radio sources II. The effects of jet magnetization}

\institute{ Dipartimento di Fisica, Universit\`a degli Studi di Torino, via
  Pietro Giuria 1, 10125 Torino, Italy \and INAF/Osservatorio Astrofisico di
  Torino, via Osservatorio 20, 10025 Pino Torinese, Italy}

\date{Received ?? / Accepted ??}

\abstract{Radio sources of low power are the most common in the   universe. Their jets typically move at nonrelativistic velocity and show  plume-like morphologies that in many instances appear distorted and bent.
  We investigate the role of magnetic field on the propagation and evolution  of low-power jets and the connection between the field intensity and the
  resulting morphology. The problem is addressed by means of three-dimensional  magnetohydrodynamic (MHD) simulations. We consider supersonic jets that
  propagate in a stratified medium. The ambient temperature increases with
  distance from the jet origin maintaining constant pressure. Jets with low
  magnetization show an enhanced collimation at small distances with respect
  to hydrodynamic (HD) cases studied in a previous paper.   These jets
  eventually evolve in a way similar to the HD cases. Jets with higher
  magnetization are affected by strong nonaxisymmetric modes that lead to the
  sudden jet energy release. From there on, distorted plumes of jet material
  move at subsonic velocities. This transition is associated with the
  formation of structures reminiscent of the ``warm spots'' observed in wide-angle-tail (WAT) sources.}


\maketitle

\section{Introduction}

The population of extended radio sources has been historically classified into
two categories \citep{FR74}, namely the Fanaroff-Riley~I (FR~I) class that
includes objects with jet-dominated emission and two-sided jets at the
kiloparsec scale smoothly extending into the external medium, and the
Fanaroff-Riley II (FR~II) class that presents lobe-dominated emission and
(often) one-sided jets at the kiloparsec scale abruptly terminating in compact
hot spots of emission.  The distorted, diffuse, and plume-like morphologies of
FR~I sources led researchers to model them as turbulent flows
\citep{Bicknell84, Bicknell86, Komissarov90a, Komissarov90b, DeYoung93}, while
the characteristics of FR~II, such as their linear structure and the hot spots
at the jet termination, are associated with hypersonic, and likely
relativistic flows.

This classification however does not fully cover the population of extended
radio sources. For instance, hybrid sources showing FR~I structure on one
side of the radio source and FR~II morphology on the other \citep{gopal00} have
been observed. Furthermore, many radio galaxies present distorted
morphologies. Various sub-classes have been defined: sources with narrow
angle tails and wide angle tails \citep{rudnick77,owen76} in which the
diffuse plumes are either smoothly or sharply bent with respect to the initial
jet direction, or the so-called S-, X-, and Z-shaped sources, where the
distortion affects the radio lobes.

The variegated morphologies of radio sources are clearly associated with the
various parameters describing their jets and the external medium in which they
expand. We can then progress in our understanding of the jet properties by
attempting to reproduce the different morphological
classes with numerical simulations. With this aim in mind, we showed for example that X-shaped sources can form 
when jets propagate into a flattened gas distribution
\citep{rossi17}. Similarly, in \cite{massaglia16} (hereinafter Paper I) we
followed the evolution of low-power jets that are able to produce
edge-darkened FR~I radio sources.

More specifically, in Paper I we explored with numerical experiments the
behavior of hydrodynamical jets with the physical parameters chosen to
reproduce the behavior of low-power jets, that is, with kinetic luminosities
$\approx 10^{42}$ ergs s$^{-1}$ or lower. We assumed a King's density
profile for the ambient medium considered in pressure equilibrium \citep{king72}.
They found
that the FR~I morphologies could be well reproduced only when the intrinsic
three dimensionality of the problem was taken into account. With a reasonable
choice of the physical parameters, it was also found that the transition
FR~I/FR~II occurs at a kinetic power of about $10^{43}$ erg~s$^{-1}$ (see also
\cite{ehlert18}).  More recently \cite{li18} explored the parameter space by means of relativistic
hydrodynamic (RHD)
simulations, varying, among others, the values of the
Lorentz $\gamma$ factor, jet-to-ambient density ratio, and Mach number. We
found, not surprisingly, that jets with low $\gamma$ are more unstable, with
the jets' head that detaches from and lags behind the bow shocks. In these cases
a radio source with FRI morphology could emerge, but, differently from
\citet{massaglia16}, it would not follow the transition to a turbulent state. In fact, one expects that jets that carry lower longitudinal momentum are more sensitive to non 
axially-symmetric unstable modes, that is, the ones that lead to the jet disruption. 

Results from Paper I have shown that FR~I morphologies emerge from low-power
and HD jets. An important ingredient that could play a significant role is the
magnetic field. Therefore, we include here the effects of magnetic field and adopt
the same numerical scheme as in Paper I. We examine the behavior of different
intensities of the field on the jet propagation by varying the plasma-$\beta$
parameter (ratio of thermal to magnetic pressure).

Three-dimensional MHD simulations of the jet--intracluster medium(ICM) interaction of high-power jets, that is, with kinetic luminosities exceeding $10^{44}$ erg~s$^{-1}$, were recently carried out by \cite{wein17}. They found that due to the high jet power the jets can reach
large distances and form low-density cavities. We here instead focused on much less powerful jets: $\sim 10^{42}$ erg~s$^{-1}$.

The plan of the paper is the following: in Sect. \ref{sec:theory} we describe
the numerical setup and show the equations we solve, in
Sect. \ref{sec:results} we present the obtained results, and in Sect.4 we summarize our findings.


\section{Numerical setup}
  \label{sec:theory}

\begin{table*}[tb] 
\begin{center} 
\caption{Parameter set used in the numerical simulations.} 
\begin{tabular}{lccclccccll} 
\hline
1 & 2 & 3 & 4 & 5\\
\hline 
  & 
$\beta$&   
$L_x \times L_y \times L_z$& 
$N_x \times N_y \times N_z$&  
Notes \\ 
\hline 
A & 10$^3$ &  $64 \times 120 \times 64$ & $512 \times 1280\times 512$ &  Near HD case    \\
B & 10$^2$ &  $64 \times 120 \times 64$ & $512 \times 1280 \times 512$ &  Very weak field case    \\
C & 10 &  $52 \times 240 \times 52$ & $320 \times 2400 \times 320$ &  Weak field case   \\ 
D &  3 &  $80 \times 200 \times 80$ & $384 \times 2000 \times 384$ &  Strong field case \\
E & 3 & $124 \times 200 \times 80$ & $480 \times 2000 \times 384$ & Strong field case + transverse wind    \\ 
\hline 
\end{tabular} 
\label{labvalues} 
\end{center} 
Column description: 1) case identifier, 2) initial plasma-$\beta$ value,  3)
domain extension, 4) number of grid points, 5) short description of the case.  
\end{table*}

\subsection{Magnetohydrodynamic equations}
  \label{sec:mhd}

The MHD equations written for the primitive variables are:
\begin{equation}\label{eq:continuity}
  \frac{\partial\rho}{\partial t} + \nabla \cdot (\rho \vec v) = 0\,,
\end{equation}
\begin{equation}
  \frac{\partial\vec v}{\partial t} + (\vec v \cdot \nabla)\vec v = - \frac{1}{\rho}\nabla P 
    + \frac{1}{\rho}\vec (\nabla \times \vec B)
 \times \vec B
     \,,
\end{equation}
\begin{equation}
  \frac{\partial P}{\partial t} + \vec v \cdot \nabla P
    + \Gamma P \nabla \cdot \vec v = 0 \,,
\end{equation}
\begin{equation}
  \frac{\partial\vec B}{\partial t} = \nabla \times (\vec v \times \vec B)
     \,,
\end{equation}
\begin{equation}\label{eq:tracer}
\frac{\partial f}{\partial t} + \vec v \cdot \nabla f = 0\,.
\end{equation}
The quantities $\rho$, $P,$ and $\vec v$ are the density, pressure, and velocity, respectively.
The magnetic field $\vec B$, which includes the factor $(4\pi)^{-1/2}$,
satisfies the condition $\nabla\cdot\vec B = 0$.
Finally,  $\Gamma = 5/3$ is the ratio of the specific heats.
The jet and external material are distinguished using a passive tracer, $f$, set equal to unity for the injected jet material and equal to zero for the ambient medium.

Equations (\ref{eq:continuity})-(\ref{eq:tracer}) were solved using the linear reconstruction of the PLUTO code \citep{PLUTO, Mignone12} and the HLLC Riemann solver. For controlling the $\nabla \cdot \vec B=0$ condition we used the constrained transport method. The equations where evolved in time using a second-order Runge-Kutta method with a Courant number fixed at 0.25.

\subsection{Initial and boundary conditions}

As in Paper I, the 3D simulations were carried out on a Cartesian domain with coordinates in the range $x\in [-L/2,L/2]$, $y\in [0,L_y]$ and $z\in [-L/2,L/2]$ (lengths are expressed in units of the jet radius; $y$ is the direction of jet propagation). At $t=0$, the domain is filled with a perfect gas at rest,  unmagnetized, with uniform pressure but spherically stratified density, according to a King-like profile \citep{king72}:
\begin{equation}
\rho(R)=\frac{1}{\eta} \frac{1}{1+\left(R/r_{\mathrm c}\right)^\alpha}
\label{eq:king}
,\end{equation}
where $R=\sqrt{x^2+y^2+z^2}$ is the spherical radius, $r_{\mathrm c}$ the core radius,  the density is measured in units of the jet density $\rho_j$,  and $\eta$ is the ratio $\rho_j / \rho_c$ between the jet density and the core density.  We set  $r_{\mathrm c}=40 r_j $ and $\alpha=2$ throughout. The ambient temperature increases similarly with radius for maintaining the pressure uniform.

We imposed zero-gradient boundary on all computational boundaries with the exception of the injection boundary located at $y=0$.  Here we prescribed inside the unit circle ($r < 1$ where $r = \sqrt{x^2 + z^2}$ is the cylindrical radius) a constant cylindrical inflow directed along the $y$ direction. The inflow jet values for density, velocity, and tracer are
\[
\rho_j = 1 \,,
\]
\[
v_{yj} = M \,,
\]
\[
f_j = 1\,,
\]
where velocity is measured in units of the jet sound speed on the axis, $c_{sj}$, and therefore $M$ represents the jet Mach number.
An azimuthal magnetic field is injected with the jet and is assumed  to result from a constant current inside $r=1$ and zero 
outside,
\begin{equation}
B_\phi = \left\{
\begin{array}{ll}
-B_m \ r & {\mathrm{for} \; } r<1, \\
-B_m/r & {\mathrm{otherwise.}}
\end{array} \right.
\end{equation}
The magnetic field strength is specified by prescribing 
the plasma-$\beta$ parameter
\begin{equation}
\beta = \frac{2 \langle{p}\rangle}{\langle{B^2_z+B_\phi^2}
\rangle}
,\end{equation}
where the averaging is in the range $[0,1]$.  The jet is 
injected in 
total pressure equilibrium, and the pressure profile is determined by the equilibrium condition
\begin{equation}
\frac{d}{dr}\left(p+\frac{B^2_z}{2}\right)=-\frac{1}{2r^2}
\frac{d\left(r^2B^2_\phi\right)}{dr} \,.
\end{equation}
Outside the jet nozzle, reflective boundary conditions hold. To avoid sharp transitions, we smoothly joined the injection and reflective boundary values in the following way:
\begin{equation} \label{eq:bc}
Q(x, z, t) = Q_r (x, z, t) + \frac{Q_j - Q_r(x,y,z,t)}{cosh\left[ ( r/r_s)^n \right]} \,,
\end{equation}
where $Q = \{\rho, v_x, \rho v_y, v_z, B_x, B_y, B_z, p, f\}$ are primitive flow variables with the exception of the jet velocity, which was replaced by the $y$-momentum. We note that $Q_r(x, z, t)$ are the corresponding time-dependent reflected values, while $Q_j$ are the constant injection values. In Eq. (\ref{eq:bc}) we set $r_s = 1$ and $n = 6$ for all variables except for the density, for which we chose $r_s = 1.4$ and $n = 8$. This choice ensured monotonicity in the first $(\rho v_y)$ and second $(\rho v_y^2)$ fluid outflow momenta \citep{Massaglia96}. We did not explicitly perturb the jet at its inlet, in contrast to \cite{Mig2010}. The growing nonaxially symmetric modes originate from numerical noise. 

The physical domain is covered by $N_x \times N_y \times N_z$
computational zones that are not necessarily uniformly spaced.
For domains with a large physical size, we employ a uniform
grid resolution in the central zones around the beam (typically for $ |x|, \ |z| \leq 6$) and a 
geometrically stretched grid elsewhere. The complete set of parameters for our simulations is given in Table \ref{labvalues}.

\subsection{Physical parameters}

\begin{figure*}[htp] 
\centering%
\subfigure{\includegraphics[width=\columnwidth]{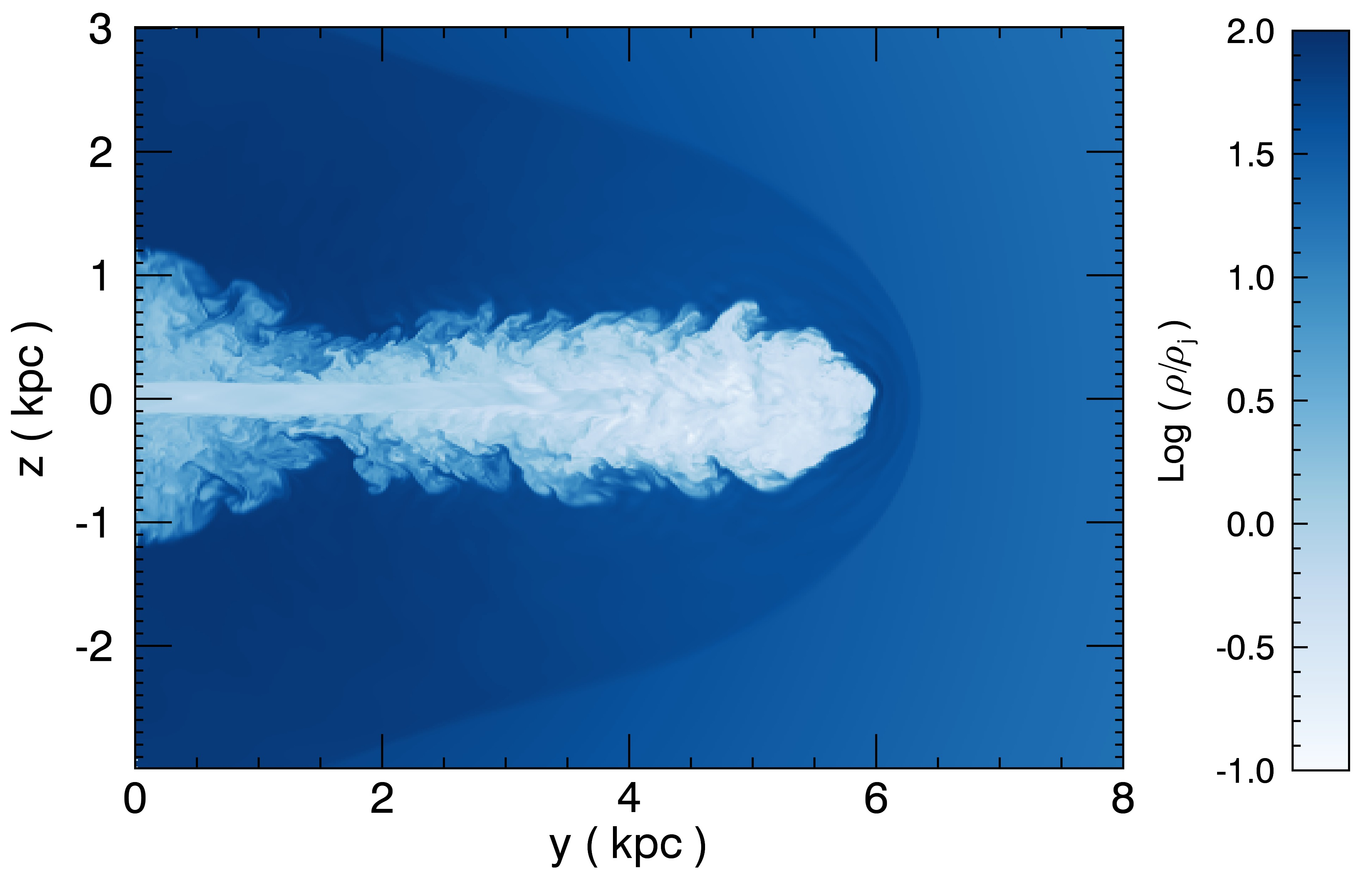}}%
\subfigure{\includegraphics[width=0.82\columnwidth]{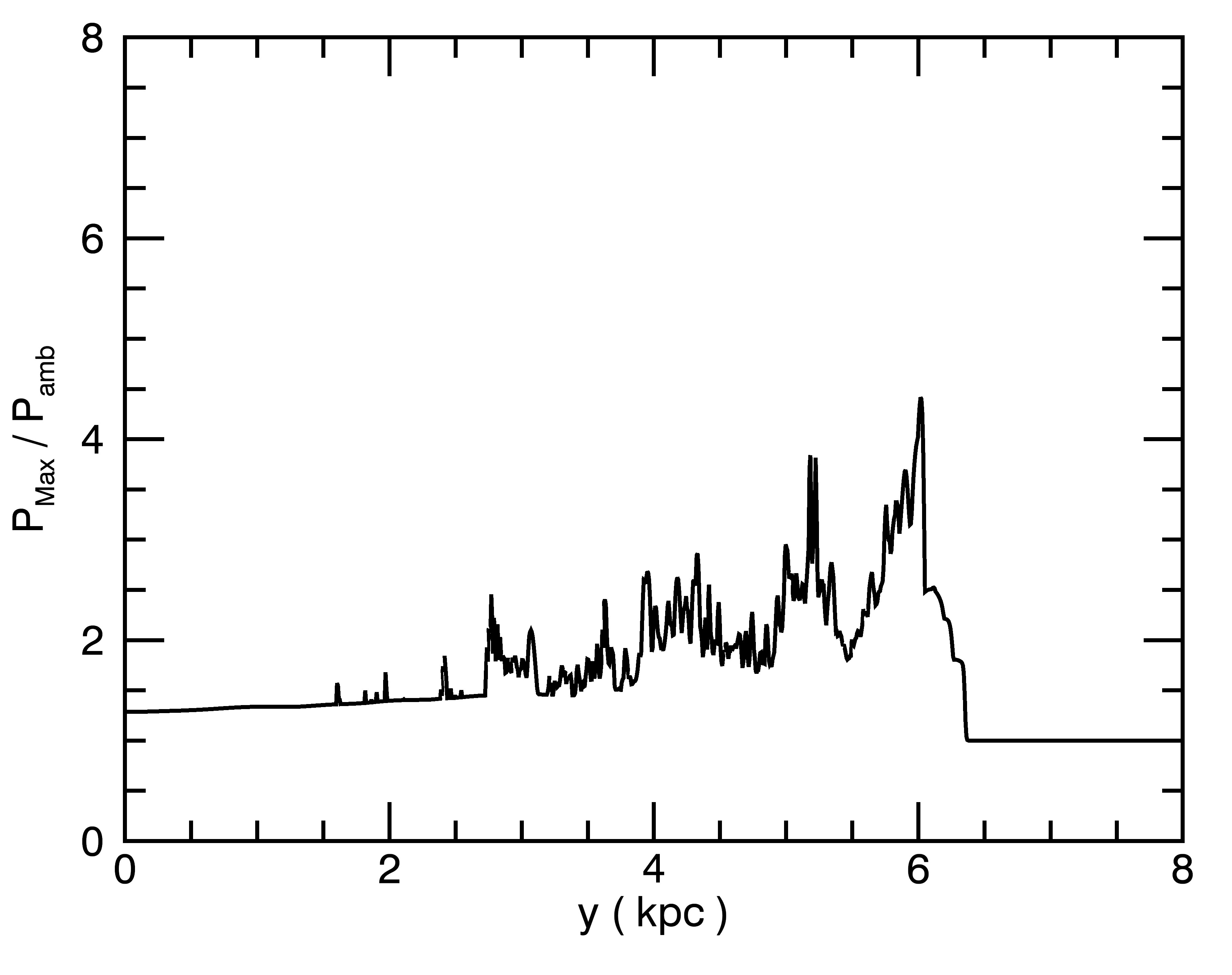}}%

\subfigure{\includegraphics[width=\columnwidth]{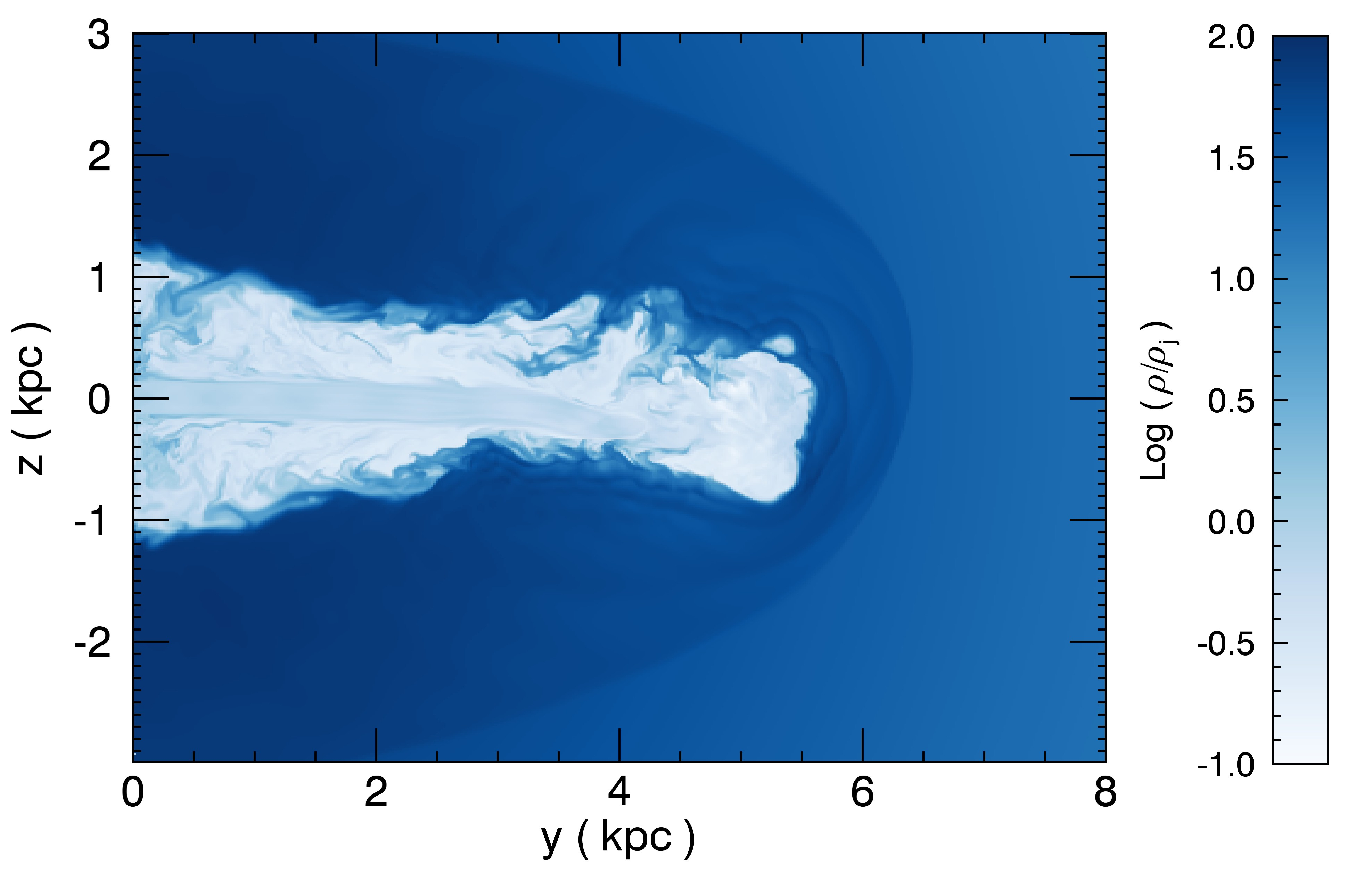}}%
\subfigure{\includegraphics[width=0.82\columnwidth]{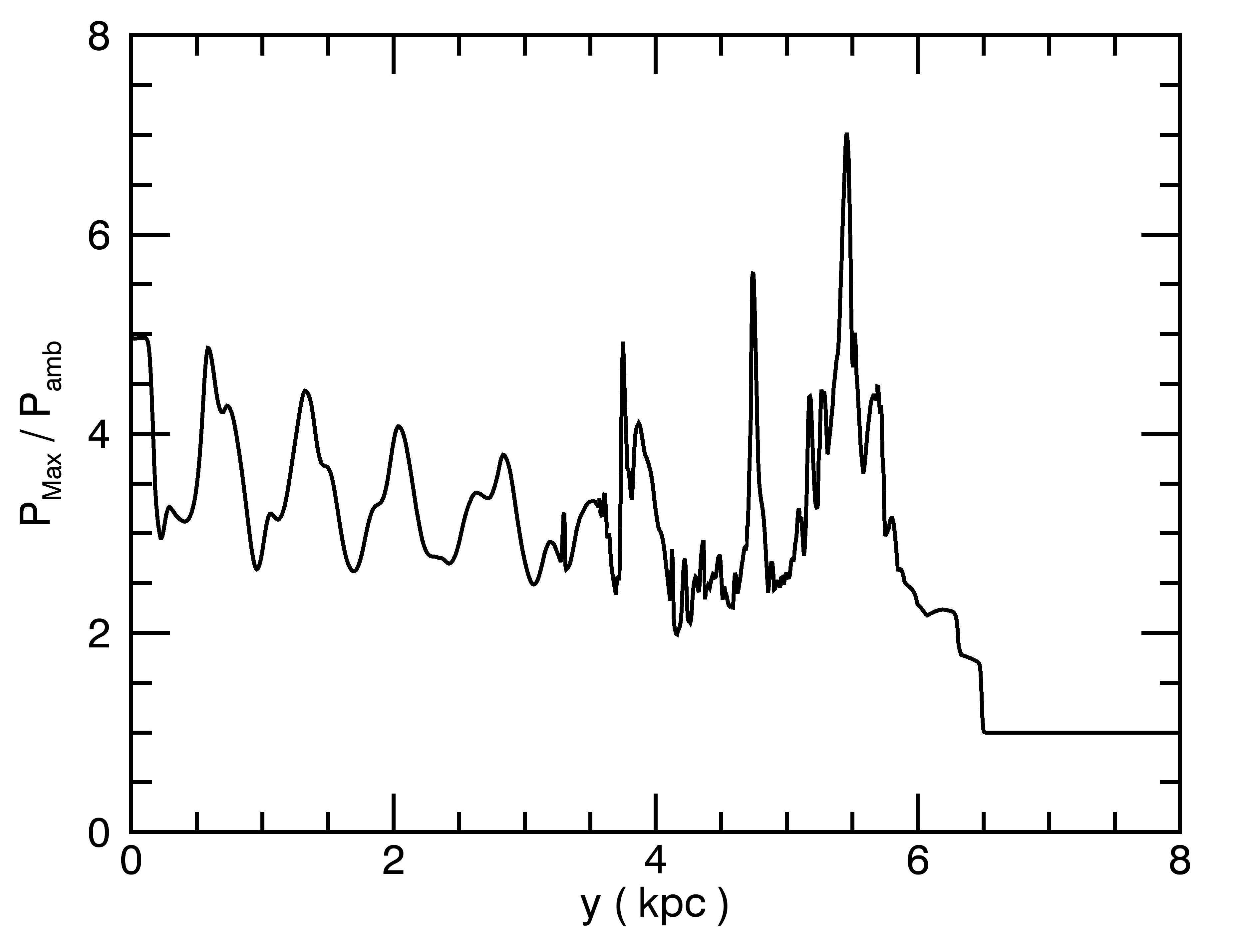}}%
\caption{Top panels: (left) Longitudinal cut of the density distribution in
  the $(y, z)$ plane for the HD case B of paper I after 200 time units, $t= 1.5 \times
  10^7$ yrs, and (right) maximum pressure as a function of $y$. Bottom panels:
  Same as in top panel but  for case D of the present paper.}
\label{fig:fr2HD} 
\end{figure*}

Since the purpose of this paper is to investigate the effects of magnetization
on the jet morphological evolution, we adopted the same parameter set of the
HD case giving a morphology more reminiscent of FR I radio sources, that is,
a density ratio $\eta = 10^{-2}$ and a Mach number $M=4$. 

For convenience, we summarize the assumptions adopted in Paper I
for connecting the numerical parameters, namely the fundamental units of
length $r_{\mathrm j}$ (jet radius), density $\rho_{\mathrm j}$ (jet density),
and velocity $c_{\mathrm sj}$ (the sound speed in the jet) with the physical
ones. We assumed a fiducial value for the jet radius of $r_{\mathrm j}=100$ pc, for
the galactic core radius $r_{\mathrm c}=4$ kpc, for the ambient medium a
central temperature of $T_{\mathrm c} = 0.2$ keV, and a particle density of
$\rho_{\mathrm c} = 1 $ cm$^{-3}$. With these assumptions, and for the cases
considered, the initial jet density results in $\rho_j = 0.01 \mathrm{cm}^{-3}$,
the jet velocity\footnote{We note that in Paper I the derived jet velocities, in physical units, are overestimated by about a factor of two. The values of all other parameter are unaffected.}$ v_j \simeq 10^9 \, M_4 \, \mathrm{cm}\, {s}^{-1} $ 
and
the jet kinetic power ${\cal L}_{\mathrm kin}= 1.1 \times 10^{42} \,
\mathrm{erg} \, \rm{s}^{-1}$.  The computational time unit $\tau$ is the sound
travel time over the initial jet radius, which in physical units corresponds
to $\tau = 7.7 \times 10^4$ yrs.

We consider matter-dominated jets, that is, with plasma-$\beta$ values always
larger than unity (see Table \ref{labvalues}), and therefore in all cases the jet
kinetic power largely exceeds the Poynting flux luminosity.  

As shown in Table \ref{labvalues}, the four cases A, B, C, and D have values of
the plasma $\beta$ equal to $10^3$, $10^2$, $10,$ and $3,$ respectively.
Concerning the actual values of the magnetic field strength, using again the
reference values discussed above, we have $B \approx 3.2 \times 10^{-6}
\,\rm{G}, 10^{-5} \,\rm{G}, 3.2 \times 10^{-5} \,\rm{G,} $ and $6 \times
10^{-5} \,\rm{G} $ for cases A, B, C, and D, respectively.  In particular the
value for case D is in good agreement with the typical estimates based on
equipartition arguments. Much smaller values of the plasma-$\beta$ would lead
to unrealistically high magnetization, very far from the equipartition value,
in many regions of the source.

\begin{figure*}[htp] 
\begin{center} 
\includegraphics[width=2\columnwidth]{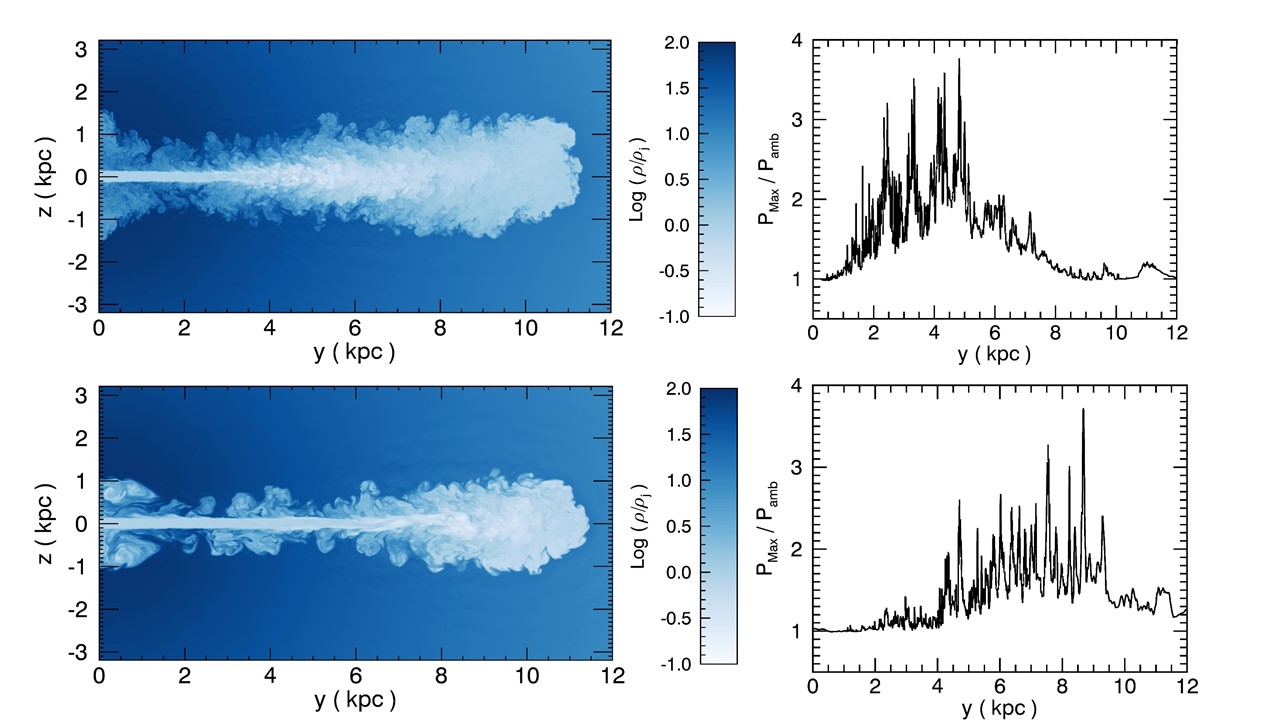}
\end{center}
\caption{Top: Results for the HD case B of Paper I ($M= 4$, $\eta = 10^{-2}$, ${\cal
    L}_{\mathrm kin} = 1.1 \times 10^{42}$) at $t=640$ time units, $4.9 \times
  10^7$ yr). Left: Cut in the $(y,z)$ plane of the logarithmic density
  distribution in units of jet density. The spatial units are in jet radii
  (100 pc) and the image extends over 12 kpc in the jet direction and over 6.4
  kpc in the transverse direction. Right: Maximum particle pressure as a
  function of $y$. The particle pressure is plotted in units of the ambient
  value (instead of the {\it computational} units $\rho_j c_{sj}^2$).  Bottom:
  As in the top panel, but for the MHD case A of the present paper and at
  $t=400$ time units, $3.1\times 10^7$ yrs. The two cases share the same jet
  parameters but the second simulation includes magnetic field corresponding
  to $\beta=10^3$.}
\label{fig:HD_B} 
\end{figure*}

\begin{figure}[htp] 
\begin{center} 
\includegraphics[width=\columnwidth]{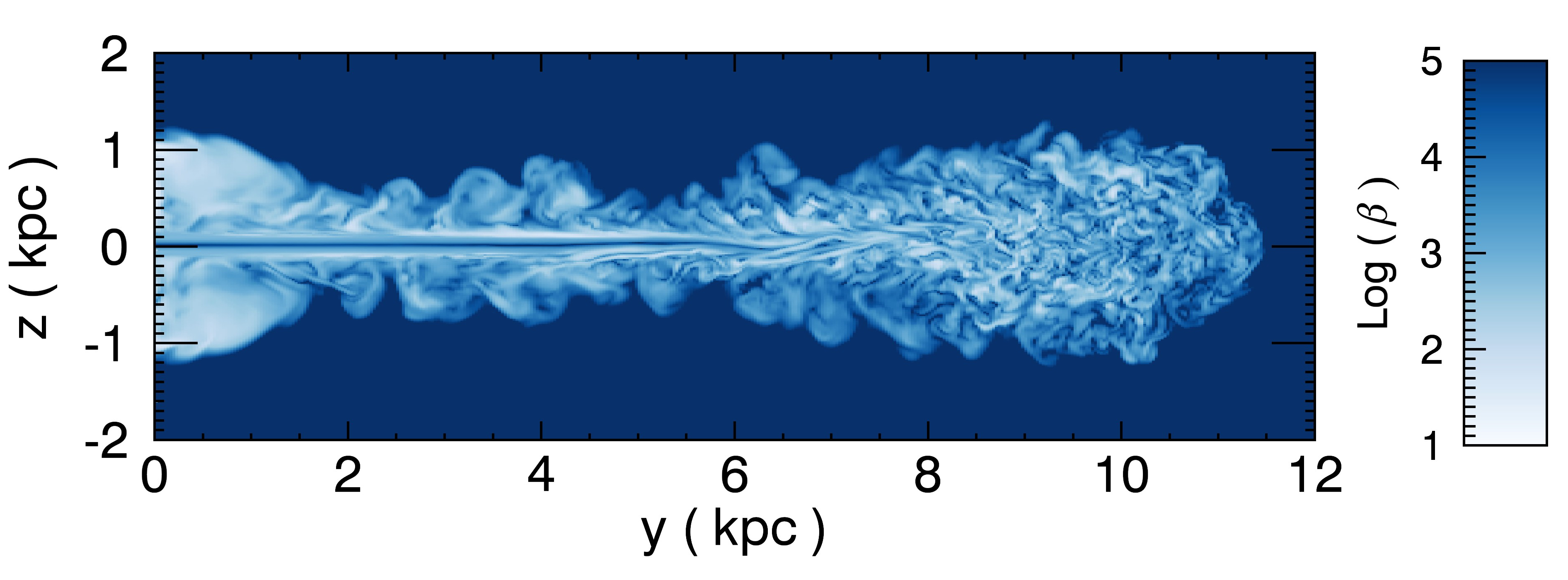}
\includegraphics[width=1.03\columnwidth]{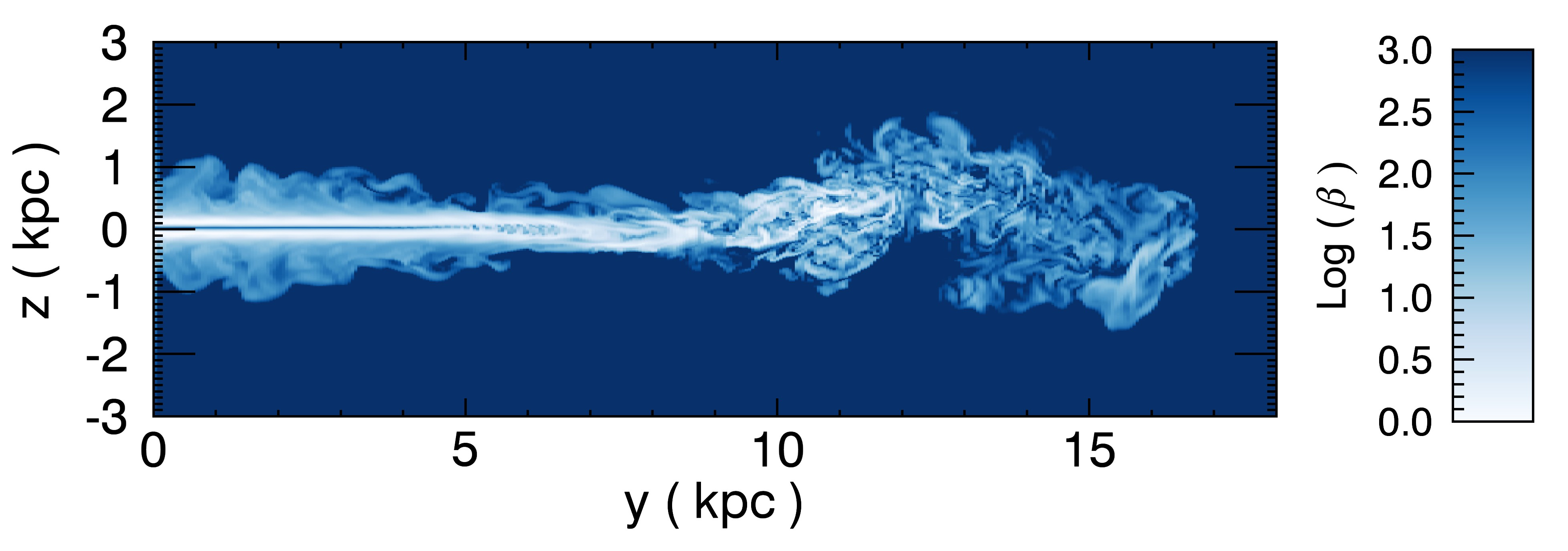}
\end{center} 
\caption{Distribution of the plasma-$\beta$ for cases A (top) and D
  (bottom).}
\label{fig:beta_A} 
\end{figure}

\section{Results}
 \label{sec:results}

\subsection{The initial stages}

It is interesting to examine the source morphology when the jet has just
exited the central galaxy core, that is, when it reaches a distance of $\sim$ 6
kpc. We consider case D of the present paper and the HD case B of Paper I, as a comparison. 

We recall that in Paper I we used the maximum pressure (i.e., the maximum pressure found at each $y$ along the jet as a function of $y$) as an
indicator of the FR~I or FR~II morphology. In the case of FR~II the
maximum pressure is found at the jet head and marks the hot spot, where the
jet energy is dissipated. For FR~I jets, conversely, dissipation occurs
gradually along the jet and the maximum pressure plot shows pressure reaching
its maximum value along the jet and then steadily decreasing.

In Fig. \ref{fig:fr2HD} we show (left panels) 
longitudinal cuts of the density distribution in the $(y,z)$ plane after 200
time units ($t= 1.5 \times 10^7$ yrs) and (right panels) 
the maximum pressure as a function of $y$. 

\begin{figure*}[htp] 
\centering%
\subfigure{\includegraphics[width=1.4\columnwidth]{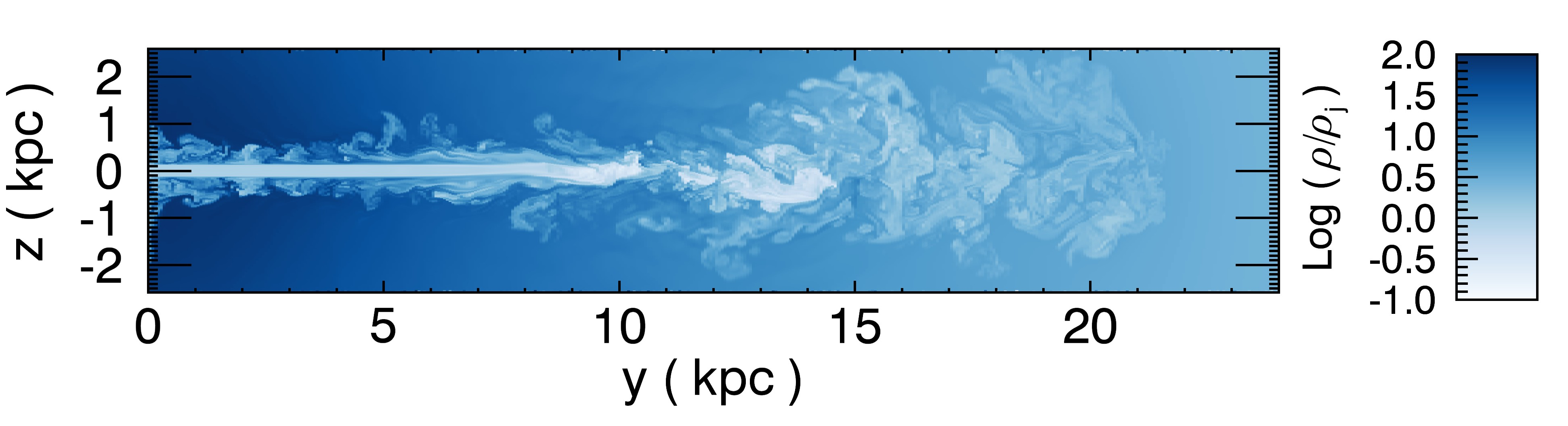}}%
\subfigure{\includegraphics[width=0.55\columnwidth]{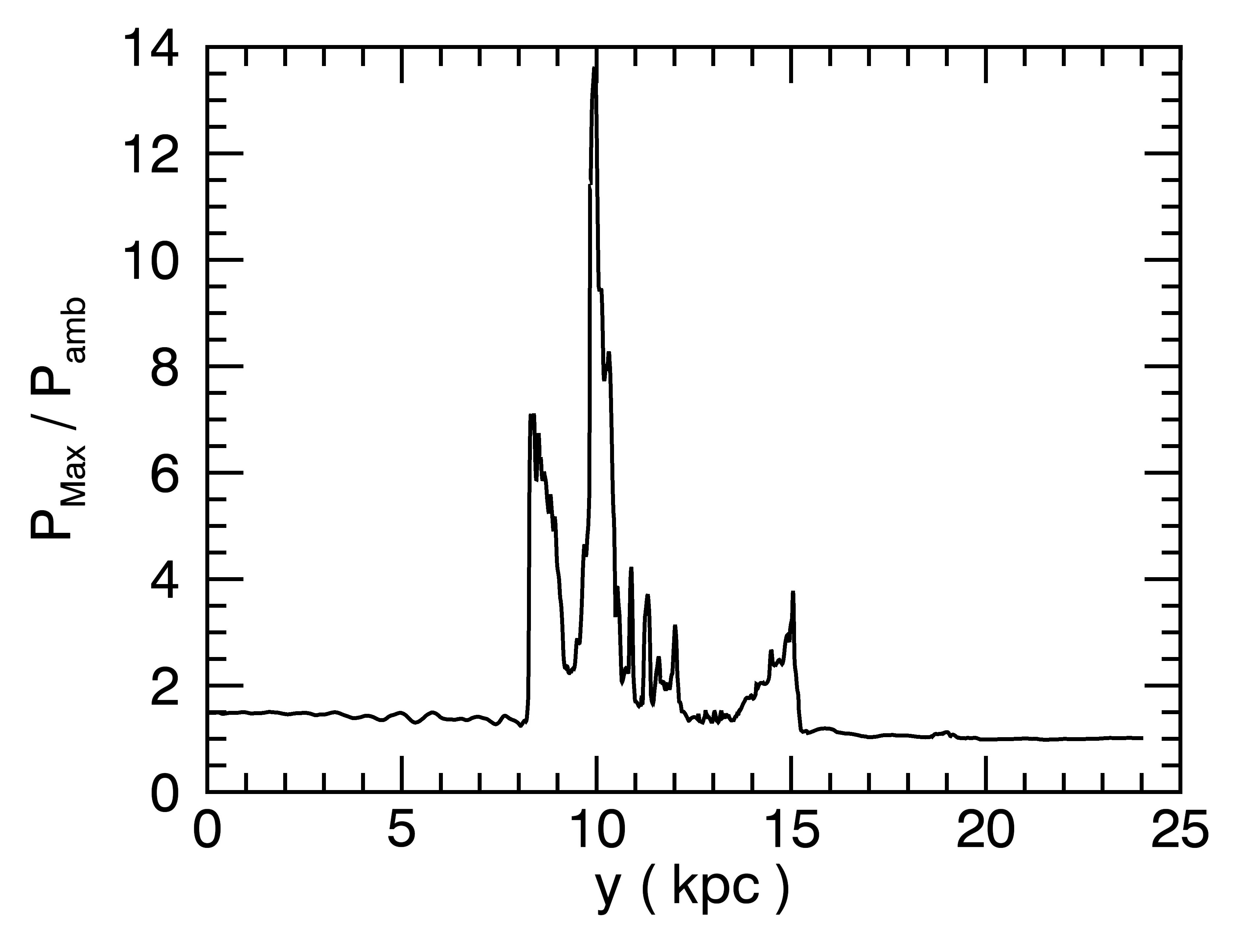}}%
\caption{Left: Cut of the logarithmic density distribution for case C in the $(y,z)$ plane; Right: Maximum particle pressure as a function of $y$ at the time $t=1100$ time units, $8.4\times 10^7$ yrs.
}
\label{fig:C} 
\end{figure*}

\begin{figure*}[htp] 
\centering%
\includegraphics[width=2\columnwidth]{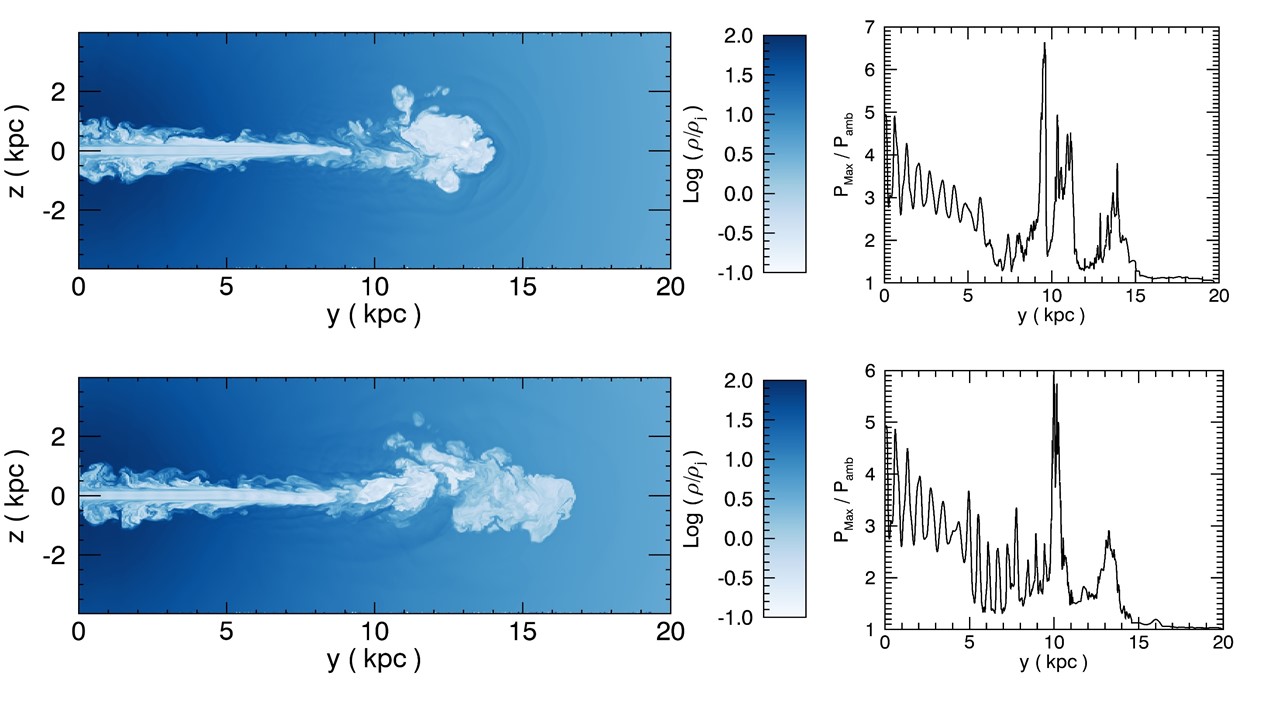}%
\caption{As in Fig. \ref{fig:C}, but for case D at $t=500$ time  units, $3.9 \times 10^7$ yrs (top panels), and $t=700$ time units, $5.4\times 10^7$ yrs (bottom panels). }
\label{fig:D} 
\end{figure*}

\begin{figure*}[htp] 
\begin{center} 
\includegraphics[width=1.8\columnwidth]{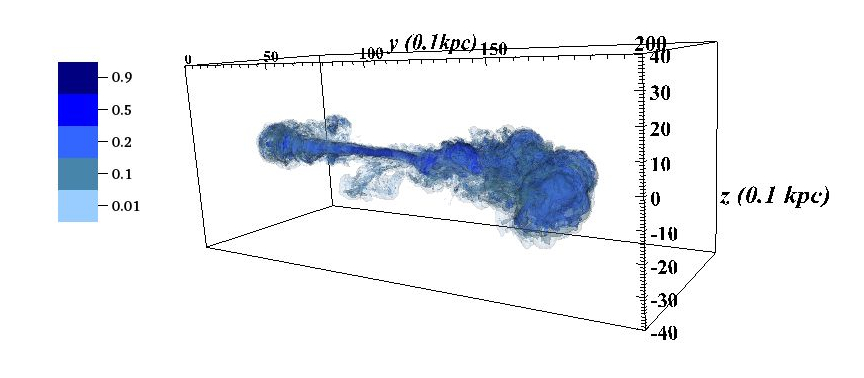} 
\end{center} 
\caption{Three-dimensional iso-contours of the tracer distribution for case D, $\beta = 3$,
  at $t=700$ time units, $5.4\times 10^7$ yrs. The size of the computational
  box is $8 \times 20 \times 8$ kpc. The $x$ axis scales like the $z$ one.}
\label{fig:3D} 
\end{figure*}

These "mini" radio sources show, in both cases, a FR~II morphology with
terminal shock at the heads od the jets; we highlight the maximum pressure behaviors and the
well-defined cocoon. In the magnetic case the nonlinear evolution of
nonaxisymmetric MHD modes begins to bend the jet, and this bending will
amplify at larger distances leading to the jet breaking and the onset of warm
spots. Furthermore, in the MHD case the inlet pressure is already larger than in the
HD case in order to maintain radial equilibrium balance. 
This pressure is then further increased in the locations of magnetic pinching.

\subsection{The effects of magnetization on the jet propagation}

We begin our analysis from case A, with $\beta=1000$, which corresponds to a
nearly negligible value of the magnetic field. We have carried out the
simulation on the same domain extension and with the same numerical resolution
as in the reference case. We would then expect similar results for the two
simulations, but comparing in Fig. \ref{fig:HD_B} the HD top-left panel with
the MHD bottom-left panel we note that while in the HD case the jet begins to
spread out after it has traveled a very short distance in the ambient medium, in
case A it maintains its collimation up to about 6 kpc before showing a
turbulent behavior. This can be explained by looking at Fig. \ref{fig:beta_A}
(top panel), where we show a cut of the plasma-$\beta$ distribution. The value
of $\beta$ at the base of the jet decreases down to $\approx 5-10$: this
implies a relevant, mainly azimuthal, contribution of the magnetic field to
the jet collimation in those regions. The increase of the field is the result
of the compression by backflow  of the jet's head in the early stages of the
propagation. The results of case B are similar to those of case A, but with the
difference that the same morphology of case A is reached at about 15-18 kpc
instead of 12 kpc.

Simulations with smaller initial values ($\beta=10-3$, cases C and D,
respectively) show that the effects of magnetization begin to be important on
the jet evolution. Stronger magnetization still has the same effect on the
initial collimation due to the enhanced decrease in $\beta$, whose minimum is
now down to 0.5 for case D (see Fig. \ref{fig:beta_A}, bottom panel). Below 10
kpc a signature of axially symmetric pinching modes is also evident.

Moreover, the azimuthal field induces nonaxially symmetric modes, as can be
seen in both cases C (Fig. \ref{fig:C}) and D (Fig. \ref{fig:D}), that cause disruption of the jet. The jet then suddenly releases most of its power at
about 10 kpc forming one or more warm spots, as revealed in the maximum
pressure plots. With respect to the classical hot spots seen in the
powerful FR~II radio galaxies, the warm spots do not represent the jet
termination point; beyond the warm spots the flow continues towards larger
radii, instead of forming a powerful backflow. In Fig. \ref{fig:D} we show
density cuts and maximum pressure at two different times ($3.9 \times 10^7$
yrs, top panels, and $5.4\times 10^7$ yrs, bottom panels) for case D: the
position, relative intensities, and structure of the warm spots are not
stationary in time, but they vary and wander between distances if $\sim$10 and $\sim$15 kpc from the origin of the jet. For case D we also show the 3D tracer
distribution in Fig. \ref{fig:3D} to show how the morphology gets distorted by the
effect of nonaxially symmetric modes.

\begin{figure}[htp] 
\centering%
\subfigure{\includegraphics[width=0.80\columnwidth]{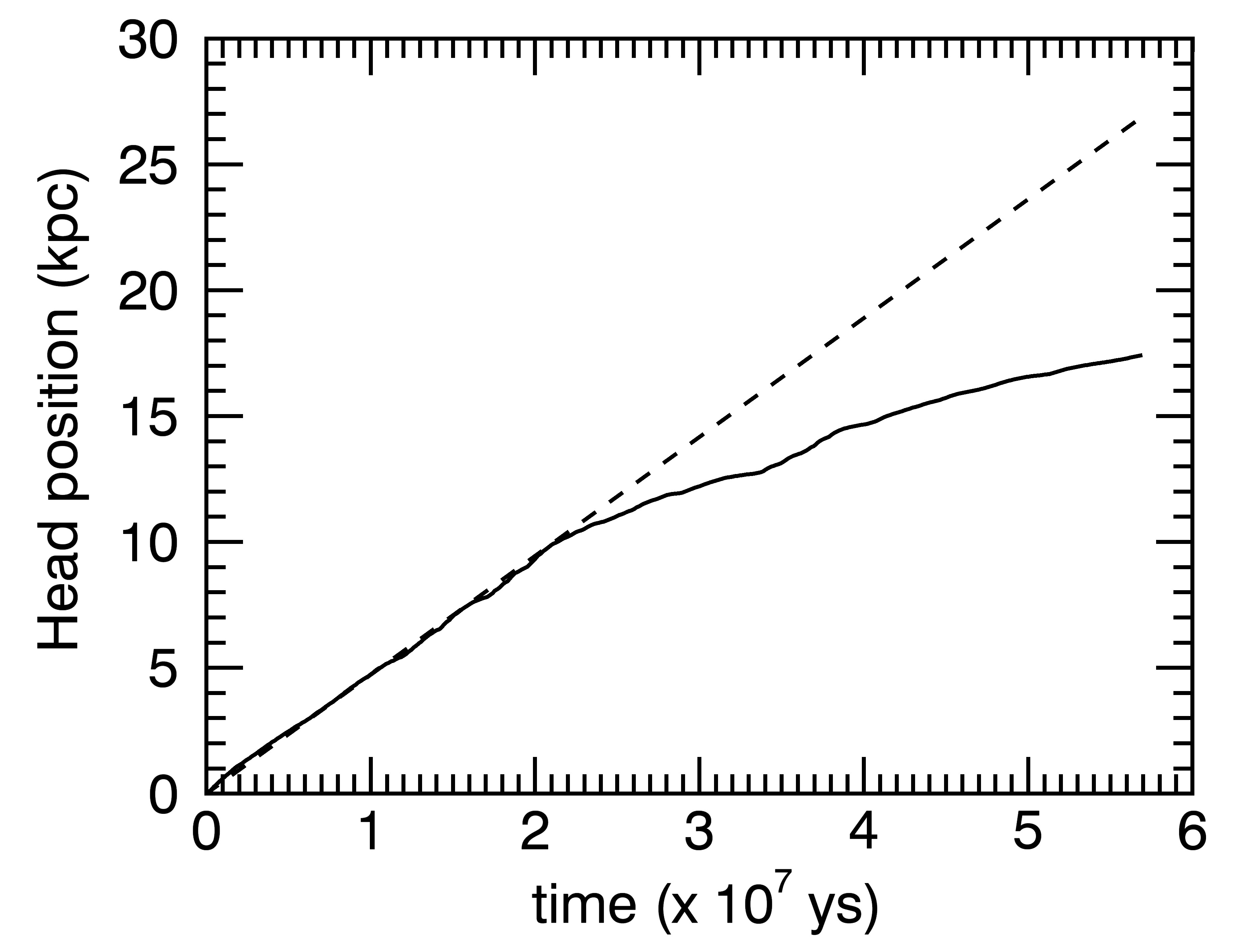}}%
\caption{Position of the head of the jet as a function of time for case D (solid line) compared to the theoretical value based on 1D momentum-conservation arguments \citep{Massaglia96}.  }
\label{fig:velhead} 
\end{figure}

Similarly to the HD simulations, in the MHD cases the advance velocity of
the head of the jet, $v_{\rm head}$, decreases with time. Figure \ref{fig:velhead}
shows the results obtained for case D: after $\sim$300 time units $v_{\rm  head}$ departs from the value based on 1D momentum conservation and this
occurs at the location of the warm spots.

\begin{figure}[htp] 
\centering%
\subfigure{\includegraphics[width=1.05\columnwidth]{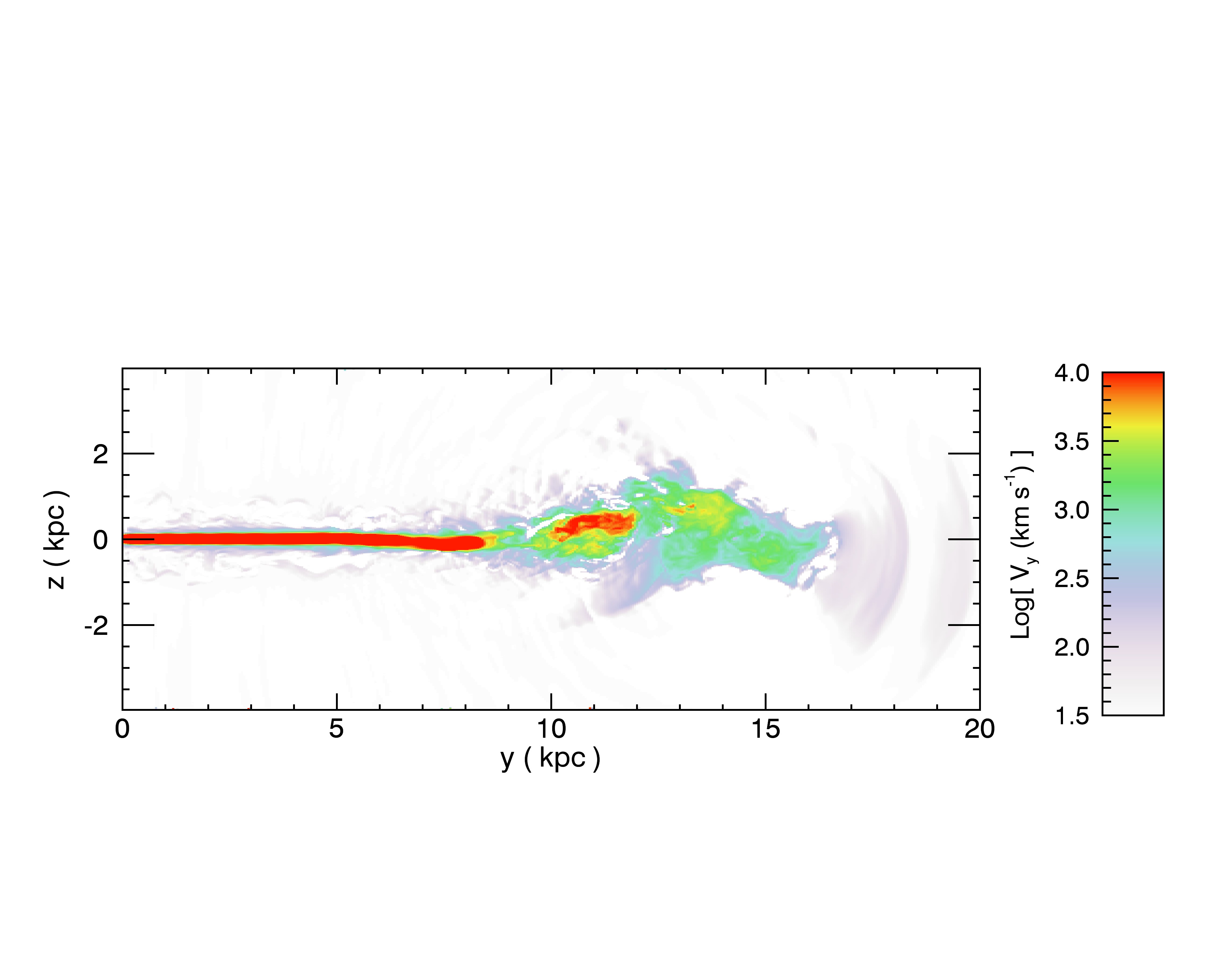}}%
\caption{Cut of the logarithm of the longitudinal velocity distribution for
    case D in the $(y,z)$ plane. We only show the positive velocities.}
\label{fig:velD} 
\end{figure}

\subsection{Distortion of the jet by a transverse wind}

In Fig. \ref{fig:velD} we see how the longitudinal velocity of the jet drops at
about 10 kpc from 10,000 km s$^{-1}$ to much smaller values: while there is
still some flow at high velocities, most of the jet material travels at 100 -
1,000 km s$^{-1}$. Therefore, one expects that the ram pressure exerted by even moderate ICM flow velocities,
with a component transverse to the jets, may be able to bend them. The
resulting morphology is likely to be reminiscent of wide-angle-tail (WAT) radio galaxies (see also \cite{soker88} for the narrow angle tail case).

We show now the results obtained by modeling the relative motion of the galaxy in the ICM by a transverse wind with constant speed $v_x$ (Case E of Table \ref{labvalues}). We assume that the intergalactic medium (IGM) acts as
a screen to the ram pressure force exerted by the ICM wind. For the transverse ICM wind velocity we adopted the form:

\begin{equation}
v_x(r_{\mathrm g} ) = v_{\infty} 
\left( 1-\frac{1} 
{1+ \left( r_{\mathrm g} /r_{\mathrm c} \right)^\alpha} \right) \, \ 
\label{eq:ICM}
,\end{equation}

where $v_{\infty} $ is the velocity of the galaxy relative to the ICM $r_{\mathrm
  g} = \sqrt{y^2+z^2}$, and $\alpha=2$. We have set $v_{\infty} =0.05$ (case
E), which in physical units corresponds to $\sim 100$ km s$^{-1}$ (see \cite{smolcic07}), consistent with the typical velocity dispersions of galaxies in groups or poor clusters \citep{wilman05}.

\begin{figure*}[htp] 
\begin{center} 
\includegraphics[width=1.8\columnwidth]{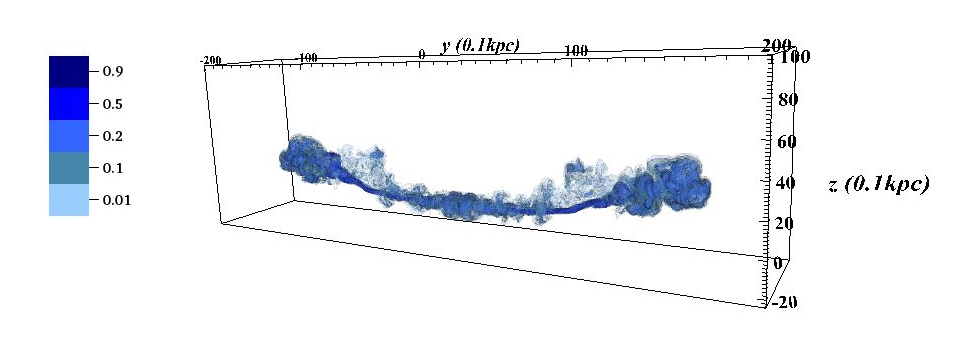} 
\end{center} 
\caption{Three-dimensional iso-contours of the tracer distribution for the WAT case E, $\beta = 3$, at $t=630$ time units, $4.8 \times 10^7$ yrs. The size of the  computational box is $12.4 \times 20 \times 8$ kpc. The $x$ axis scales as in Fig. \ref{fig:3D}.}

\label{fig:3DWAT} 
\end{figure*}

\begin{figure*}
\begin{center} 
\includegraphics[width=2\columnwidth]{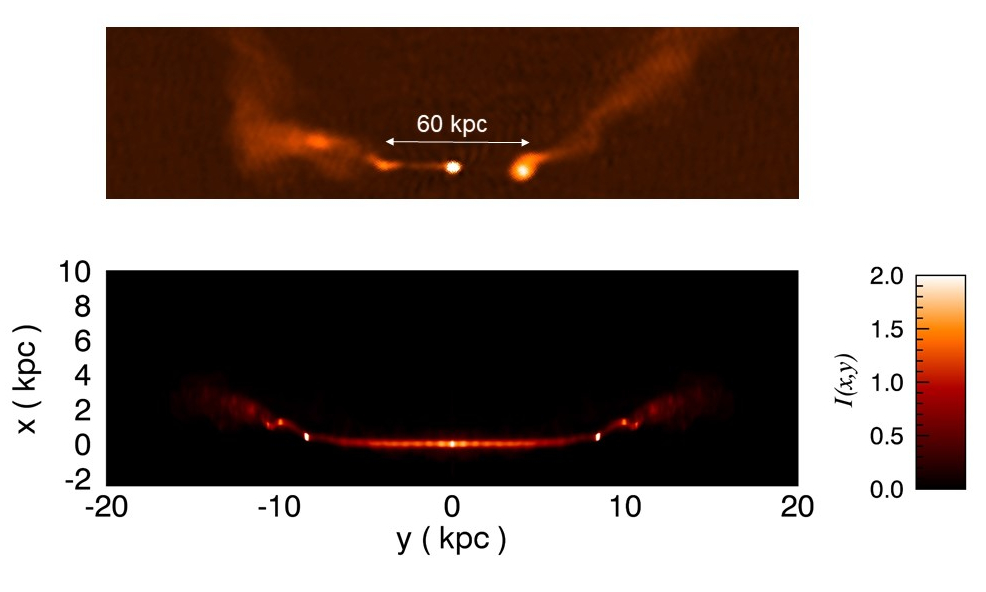} 
\end{center} 
\caption{Top: Radio image at 4.6 GHz of 3C~465, obtained from the NRAO VLA
  Archive Survey Images Page (and rotated counterclockwise by 145$^\circ$)
  showing the typical morphology of WATs. The resolution is $4\arcsec$ and the
  FOV is $2\arcmin \times 7\arcmin$, corresponding to $70\times 250$
  kpc. Bottom: Cut in the $(y,z)$ plane of the (smoothed) brightness 
  distribution in arbitrary units at $t=630$ time units for case E.  }
\label{3c465} 
\end{figure*}

In Fig. \ref{fig:3DWAT} we show the iso-contours of the tracer distribution
for case E at $t=630$ time units, $t=4.8 \times 10^7$ yrs. We note the expected
bending of the jets at a distance of about 8-9 kpc from the origin, i.e., where
they disrupt. In Fig. \ref{3c465} we compare the radio image of a WAT source (namely, 3C~465) with a synthetic brightness distribution ${\mathcal{I}}(x,y)$ map for case E. This is obtained by integrating along the line-of-sight the 3D distribution of emissivity $\epsilon(x,y,z)$, assumed to be proportional to $P(x, y, z)  B^2(x, y, z)$:

\begin{equation}
{\mathcal{I}}(x,y) = \int_{-z_{\mathrm{max}}}^{z_{\mathrm{max}}} \epsilon(x,y,z) \ dz \, .
\label{eq:bright}
\end{equation}

\begin{figure}[htp] 
\centering%
\subfigure{\includegraphics[width=\columnwidth]{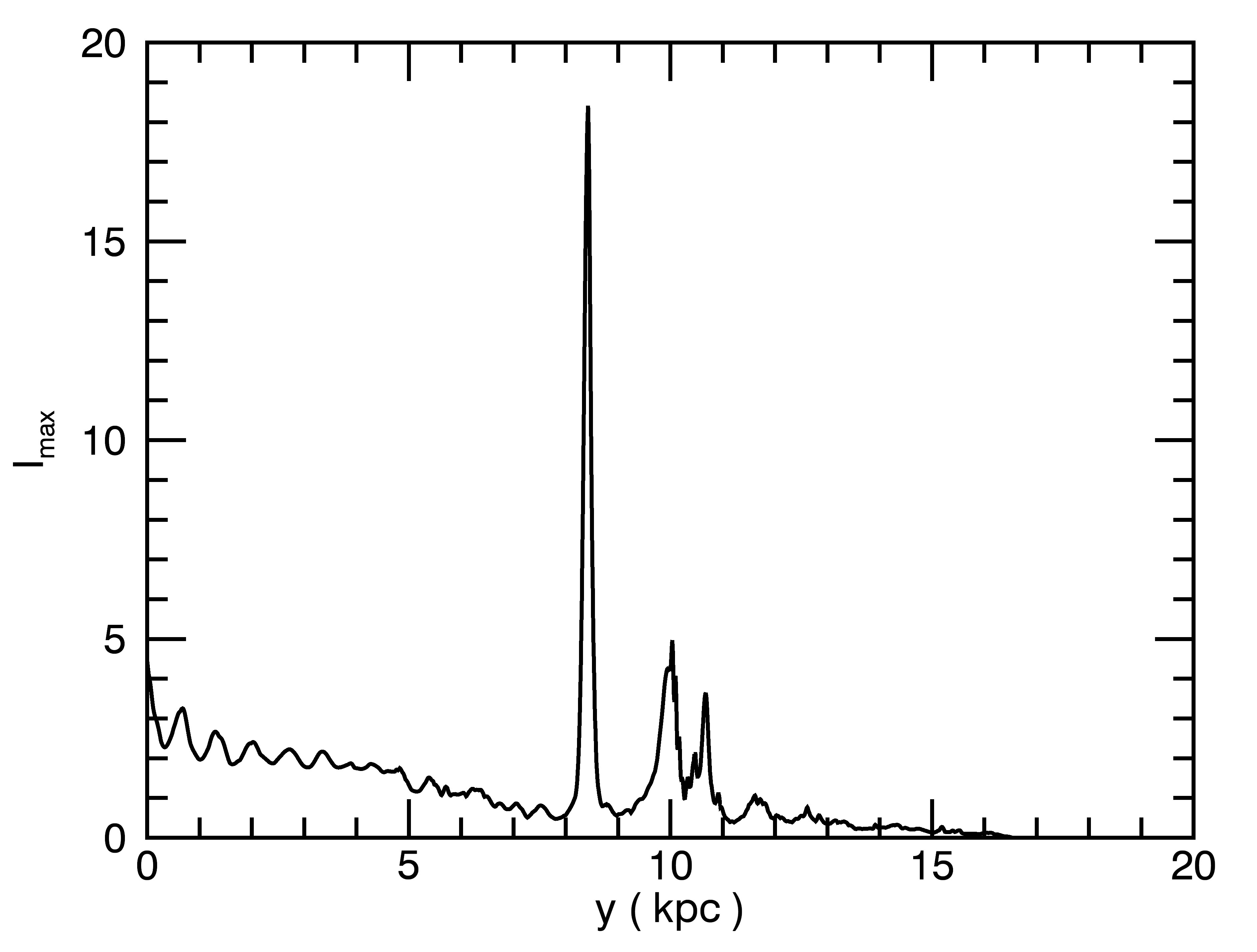}}%
\caption{Maximum brightness as a function of $y$ at the time $t=630$ time units for case E.}
\label{fig:Wpmax} 
\end{figure}

 We note the remarkable similarity
of these two images but also the rather different linear scales. In the synthetic map we note the presence of a warm spot at $\sim 8-9$ kpc, which can be clearly seen in the plot of the maximum brightness in Fig. \ref{fig:Wpmax}.

\section{Summary and conclusions}
  \label{sec:conclusion}

In the present paper we focused our attention on the effects of the
magnetization of low-power jets. The main outcomes of this investigation can
be summarized as follows:

\begin{itemize}
\item Large values of plasma-$\beta$ ($\sim$ 100 $-$ 1000) have the main
  effect of providing an additional initial collimation to the jet. In the
  early stages of the propagation the effective $\beta$ decreases to
  $\beta \sim 5-10$ due to the compression by the backflow. Since the
  magnetic field is mainly azimuthal the jet collimation is favored. The
  later evolution is similar to the HD case leading to typical FR~I
  morphologies.

\item Smaller values of the plasma parameter ($\beta \sim 3 - 10$) have the
  same effect on the initial collimation due to an even stronger decrease in
  the value of $\beta$. At later stages, strong nonaxially symmetric unstable
  modes develop leading to the sudden jet disruption at a distance of $\sim 10
  - 15 $ kpc. Here, the jet releases most of its power, forming regions of
  strong emission ("warm spots") as revealed by the maximum pressure
  plots. From there on, the collimation is lost and the fluid advances at
  subsonic speed forming diffuse plumes. This is what distinguishes these flows
  from those of the FR~II radio galaxies, where the hot spots are
  located at the termination point of the jet. The warm spots are highly dynamic
  structures, rapidly changing with time, but always located within a small
  range of distances. 

\item The abrupt velocity drop of the flow can lead, in   the presence of a transverse flow (such as that due to the motion of the host galaxy), to   a bending of the jet beyond the warm spots. This produces a WAT-like   morphology.

\end{itemize}

The analysis of the properties of these low-power jets indicates that in both
the HD and MHD cases the radio galaxies in their initial phases present the
peak of maximum pressure at the head of the jet, a feature typical of FR~IIs, at
distances reaching about 6-8 kpc. However, radio sources with an FR~I
morphology and sizes in the range of only a few kiloparsecs are often observed
\citep{balmaverde06}. We showed that the presence of magnetic field in these
phases actually increases the jet collimation with respect to the HD
case. Therefore the small FR~Is are likely to be associated with jets of even 
lower kinetic power than those examined here (${\cal L}_{\mathrm kin} \sim 10^{42} \,
\mathrm{erg} \, \rm{s}^{-1}$), in which the effects of turbulent onset destroy the jet collimation on these small scales.

The MHD simulations produce morphology, with a rapid transition from a
collimated jet terminating into warm spots, to diffuse plumes
reminiscent of WAT radio galaxies. Observations of WATs indicate that the
separation between the two warm spots spans a large range of values, from
$R_{\rm WS} \sim 30$ to $\sim 240$ kpc (e.g., \citealt{odonoghue93}). In our
simulations they are located at $R_{\rm WS} \sim 10 - 15$ kpc and are quasi
stationary, that is, their distance from the nucleus does not steadily increase
with time. The largest values of $R_{\rm WS}$ appear difficult to be
reconciled with the simulations presented here. Most likely, larger WATs can be
produced by increasing the jet power. Our simulations are performed with a
rather low jet power, ${\cal L}_{\mathrm kin}= 1.1 \times 10^{42} \,
\mathrm{erg} \, \rm{s}^{-1}$. In Paper I we estimated that the transition
between the two Fanaroff-Riley classes occurs at a jet power about one order
of magnitude higher, while the radio power of WATs straddles the separation
between FR~I and FR~II \citep{owen76}. Furthermore, the jets in WATs are often
highly asymmetric at a distance of several kiloparsecs from the nucleus
\citep{odonoghue93}, an indication of relativistic motions.

Another characteristic feature of WATs is that the diffuse plumes are bent
with respect to the initial jet direction. Although the actual mechanism for
bending the WAT tails remains unknown and under debate (e.g.,
\citealt{sakelliou00}) we showed that a relative velocity of about 100 km
s$^{-1}$ is sufficient to bend the flow after the warm spots because the flow
velocity drops suddenly.

In this paper we focused on only one parameter set, with fixed values of the
Mach number and jet power, and only varied the $\beta$ parameter. A more
detailed investigation of the parameter space is certainly required. In
particular, as noted above, WATs are likely sources associated with
relativistic jets of higher power with respect to those studied in these
simulations.

In this respect, it is worth noting that the high $\beta$ value (and the HD
case) seems to indicate that an FR~I morphology is obtained with a low value
of the magnetic field strength. However, this is not necessarily the case
because $\beta$ is the ratio between the thermal and magnetic pressure. At
fixed $B$, one can obtain a high value of $\beta$ by increasing the internal pressure of the jet. This results in a decrease of the Mach number (because $M
\propto P^{-1}$) and therefore FR~Is might have larger $\beta$ as a consequence of
a lower Mach number with respect to WATs. This is in line with the idea that the
low Mach number flows are more prone to turbulence.

In Paper I and in the present paper we have used   numerical
simulations to analyze
the dynamics of low-power jets as they propagate within the ISM at first, and
then emerge into the ICM. On the bases of the tracer and density distributions
and on the pressure behavior we have inferred the appearance of these objects
in radio observations. A more realistic approach, however, would be to follow, alongside
the jet dynamics, the evolution of the population of  relativistic electrons from which 
the nonthermal radiation originates.
Some steps in this direction have been performed by \citet{fromm16} and \citet{turner18} and a more refined treatment has recently
been proposed by \citet{vaidya18} as  a new particle module implemented in PLUTO. 
Further simulations of the low-power jets considered in this paper, using this module  and producing more realistic synthetic radio brightness and polarization maps, will be the subject of a forthcoming paper.

\begin{acknowledgements}
 We acknowledge  support by ISCRA and by the accordo quadro INAF-CINECA 2017 for the availability of high-performance computing resources. We acknowledge also support from PRIN MIUR 2015 (grant number 2015L5EE2Y) .
 \end{acknowledgements}

\bibliographystyle{aa}
\bibliography{./paper_MHD.bib}

\begin{thebibliography}{28}
\expandafter\ifx\csname natexlab\endcsname\relax\def\natexlab#1{#1}\fi

\bibitem[{{Balmaverde} \& {Capetti}(2006)}]{balmaverde06}
{Balmaverde}, B. \& {Capetti}, A. 2006, \aap, 447, 97

\bibitem[{{Bicknell}(1984)}]{Bicknell84}
{Bicknell}, G.~V. 1984, \apj, 286, 68

\bibitem[{{Bicknell}(1986)}]{Bicknell86}
{Bicknell}, G.~V. 1986, \apj, 300, 591

\bibitem[{{De Young}(1993)}]{DeYoung93}
{De Young}, D.~S. 1993, \apjl, 405, L13

\bibitem[{{Ehlert} {et~al.}(2018){Ehlert}, {Weinberger}, {Pfrommer}, {Pakmor},
  \& {Springel}}]{ehlert18}
{Ehlert}, C., {Weinberger}, R., {Pfrommer}, C., {Pakmor}, R., \& {Springel}, V.
  2018, {arXiv:1806.05679v1}

\bibitem[{{Fanaroff} \& {Riley}(1974)}]{FR74}
{Fanaroff}, B.~L. \& {Riley}, J.~M. 1974, \mnras, 167, 31P

\bibitem[{{Fromm} {et~al.}(2016){Fromm}, {Perucho}, {Mimica}, \&
  {Ros}}]{fromm16}
{Fromm}, C.~M., {Perucho}, M., {Mimica}, P., \& {Ros}, E. 2016, \aap, 588, A101

\bibitem[{{Gopal-Krishna} \& {Wiita}(2000)}]{gopal00}
{Gopal-Krishna} \& {Wiita}, P.~J. 2000, \aap, 363, 507

\bibitem[{{King}(1972)}]{king72}
{King}, I.~R. 1972, \apj, 174, L123

\bibitem[{{Komissarov}(1990{\natexlab{a}})}]{Komissarov90a}
{Komissarov}, S.~S. 1990{\natexlab{a}}, \apss, 165, 313

\bibitem[{{Komissarov}(1990{\natexlab{b}})}]{Komissarov90b}
{Komissarov}, S.~S. 1990{\natexlab{b}}, \apss, 165, 325

\bibitem[{{Li} {et~al.}(2018){Li}, {Wiita}, {Schuh}, {Elghossain}, \&
  {Hu}}]{li18}
{Li}, Y., {Wiita}, P.~J., {Schuh}, T., {Elghossain}, G., \& {Hu}, S. 2018,
  {arXiv:1810.00515v1}

\bibitem[{{Massaglia} {et~al.}(1996){Massaglia}, {Bodo}, \&
  {Ferrari}}]{Massaglia96}
{Massaglia}, S., {Bodo}, G., \& {Ferrari}, A. 1996, \aap, 307, 997

\bibitem[{{Massaglia} {et~al.}(2016){Massaglia}, {Bodo}, {Rossi}, {Capetti}, \&
  {Mignone}}]{massaglia16}
{Massaglia}, S., {Bodo}, G., {Rossi}, P., {Capetti}, S., \& {Mignone}, A. 2016,
  \aap, 596, A12

\bibitem[{{Mignone} {et~al.}(2007){Mignone}, {Bodo}, {Massaglia}, {Matsakos},
  {Tesileanu}, {Zanni}, \& {Ferrari}}]{PLUTO}
{Mignone}, A., {Bodo}, G., {Massaglia}, S., {et~al.} 2007, \apjs, 170, 228

\bibitem[{{Mignone} {et~al.}(2010){Mignone}, {Rossi}, {Bodo}, {Ferrari}, \&
  {Massaglia}}]{Mig2010}
{Mignone}, A., {Rossi}, P., {Bodo}, G., {Ferrari}, A., \& {Massaglia}, S. 2010,
  \mnras, 402, 7

\bibitem[{{Mignone} {et~al.}(2012){Mignone}, {Zanni}, {Tzeferacos}, {van
  Straalen}, {Colella}, \& {Bodo}}]{Mignone12}
{Mignone}, A., {Zanni}, C., {Tzeferacos}, P., {et~al.} 2012, \apjs, 198, 7

\bibitem[{{O'Donoghue} {et~al.}(1993){O'Donoghue}, {Eilek}, \&
  {Owen}}]{odonoghue93}
{O'Donoghue}, A.~A., {Eilek}, J.~A., \& {Owen}, F.~N. 1993, \apj, 408, 428

\bibitem[{{Owen} \& {Rudnick}(1976)}]{owen76}
{Owen}, F.~N. \& {Rudnick}, L. 1976, \apjl, 205, L1

\bibitem[{{Rossi} {et~al.}(2017){Rossi}, {Bodo}, {Capetti}, \&
  {Massaglia}}]{rossi17}
{Rossi}, P., {Bodo}, G., {Capetti}, S., \& {Massaglia}, S. 2017, \aap, 606, A57

\bibitem[{{Rudnick} \& {Owen}(1977)}]{rudnick77}
{Rudnick}, L. \& {Owen}, F.~N. 1977, \aj, 82, 1

\bibitem[{{Sakelliou} \& {Merrifield}(2000)}]{sakelliou00}
{Sakelliou}, I. \& {Merrifield}, M.~R. 2000, \mnras, 311, 649

\bibitem[{{Smol\v{c}i\'{c}} {et~al.}(2007){Smol\v{c}i\'{c}}, {Schinnerer},
  {Finoguenov}, {Sakelliou}, {Carilli}, {Botzler}, {Brusa}, {Scoville},
  {Ajiki}, {Capak}, {Guzzo}, {Hasinger}, {Impey}, {Jahnke}, {Kartaltepe},
  {McCracken}, {Mobasher}, {Murayama}, {Sasaki}, {Shioya}, {Taniguchi}, \&
  {Trump}}]{smolcic07}
{Smol\v{c}i\'{c}}, V., {Schinnerer}, E., {Finoguenov}, A., {et~al.} 2007,
  \apjs, 172, 295

\bibitem[{{Soker} {et~al.}(1988){Soker}, {O'Dea}, \& {Sarazin}}]{soker88}
{Soker}, N., {O'Dea}, C.~P., \& {Sarazin}, C.~L. 1988, \apj, 327, 627

\bibitem[{{Turner} {et~al.}(2018){Turner}, {Rogers}, {Shabala}, \&
  {Krause}}]{turner18}
{Turner}, R.~J., {Rogers}, J.~G., {Shabala}, S.~S., \& {Krause}, M. G.~H. 2018,
  \mnras, 473, 4179

\bibitem[{{Vaidya} {et~al.}(2018){Vaidya}, {Mignone}, {Bodo}, {Rossi}, \&
  {Massaglia}}]{vaidya18}
{Vaidya}, B., {Mignone}, A., {Bodo}, G., {Rossi}, P., \& {Massaglia}, S. 2018,
  \apj, 865, 144

\bibitem[{{Weinberger} {et~al.}(2017){Weinberger}, {Ehlert}, {Pfrommer},
  {Pakmor}, \& {Springel}}]{wein17}
{Weinberger}, R., {Ehlert}, C., {Pfrommer}, C., {Pakmor}, R., \& {Springel}, V.
  2017, \mnras, 470, 4530

\bibitem[{{Wilman} {et~al.}(2005){Wilman}, {Balogh}, {Mulchaey}, {Oemler},
  {Carlberg}, \& {Morris}}]{wilman05}
{Wilman}, D.~J., {Balogh}, M.~L., {Mulchaey}, J.~S., {et~al.} 2005, \mnras,
  358, 71

\end{thebibliography}

\end{document}